\documentclass[twocolumn,dvipsnames]{aastex62}

\usepackage{amsmath}
\usepackage{amssymb}
\usepackage{bm}
\usepackage[normalem]{ulem}

\newcommand{\boldell}{\bm{\ell}}

\begin{document}
\title{The Power Spectra of Polarized, Dusty Filaments}
\author{Kevin M. Huffenberger}
\affiliation{Department of Physics, Florida State University, Tallahassee, Florida 32306, USA}

\author{Aditya Rotti}
\affiliation{Jodrell Bank Centre for Astrophysics, School of Physics and Astronomy, The University of Manchester, Manchester M13 9PL, UK}

\author{David C. Collins}
\affiliation{Department of Physics, Florida State University, Tallahassee, Florida 32306, USA}

 
\begin{abstract}
  
We develop an analytic model for the power spectra of polarized filamentary structures as a way to study the Galactic polarization foreground to the Cosmic Microwave Background.   Our approach is akin to the cosmological halo-model framework, and reproduces the main features of the Planck 353 GHz power spectra.   
We model the foreground as randomly-oriented, three-dimensional, spheroidal filaments, accounting for their projection onto the sky. 
The main tunable parameters are the distribution of filament sizes, the filament physical aspect ratio, and the dispersion of the filament axis around the local magnetic field direction.
The abundance and properties of filaments as a function of size determine the slopes of the foreground power spectra, as we show via scaling arguments.  The filament aspect ratio determines the ratio of $B$-mode power to $E$-mode power, and specifically reproduces the Planck-observed dust ratio of one-half when the short axis is roughly one-fourth the length of the long axis.  Filament misalignment to the local magnetic field determines the $TE$ cross-correlation, and to reproduce Planck measurements, we need a (three-dimensional) misalignment angle with a root mean squared dispersion of about 50 degrees.  
These parameters are not sensitive to the particular filament density profile.
By artificially skewing the distribution of the misalignment angle, this model can reproduce the Planck-observed (and parity-violating) $TB$ correlation.  The skewing of the misalignment angle necessary to explain $TB$ will cause a yet-unobserved, positive $EB$ dust correlation, a possible target for future experiments.

\end{abstract}


\section{Introduction}
Polarized Galactic microwave emission poses a challenge to the search for
primordial $B$-mode polarization in the Cosmic Microwave Background (CMB).  The
$B$-mode signal could provide direct constraints on the energy scale of
inflation, but the Milky Way foregrounds may outshine it at all frequencies, everywhere on the sky \citep{2016A&A...586A.133P,2016arXiv161002743A, 2016MNRAS.462.2063H, 2017MNRAS.469.2821T, 2018JCAP...04..023R,2019arXiv190210541H}.
We have a limited understanding of this foreground, and we must learn more to ensure the reliability of future $B$-mode measurements.

The foreground emission involves the turbulent interplay of gas, dust, and magnetic fields in the Galaxy's interstellar medium (ISM).  The magnetic fields organize the flow and control the precession of grains that give rise to the polarized dust signal.  Although Planck's 353 GHz polarization channel has given us a first look, several features of the dust polarization remain without physical explanations.

For example, the amplitude of dust polarization $B$-mode power is approximately half of $E$-mode power,  $A_{BB}/A_{EE} = 0.53 \pm 0.01$, when fit on a large portion of the sky (for $f_{\rm  sky}^{\rm eff}=0.52$--$0.71$).
Smaller patches also show the same mean value $ A_{BB}/A_{EE} = 0.51$, with small patch-to-patch dispersion $\sigma_{BB/EE} = 0.18$ \citep{2016A&A...586A.133P,2018arXiv180104945P}. This observation defied pre-Planck expectations.  Random polarization orientations, or coherent orientations overlaying random polarization intensity fluctuations, both yield equal amounts of $E$ and $B$ \citep{2001PhRvD..64j3001Z,2014PhRvL.113s1303K}.

We have some understanding of dust physics and its relationship to polarization modes.
The amplitude and orientation of the dust signal is set by the integrated
column density and magnetic field orientation.
For $E$ to have more power than $B$
qualitatively means that density fluctuations (structures in the ISM density
field) must prefer orientations parallel or perpendicular to the local magnetic
field \citep{2018arXiv180711940R}.
This picture is borne out by measurements of the magnetic field orientation in individual, bright, filamentary structures in the Planck 353 GHz data \citep{2016A&A...586A.141P}.  This is further validated by the observations that linear structures in neutral hydrogen emission, highlighted by a Rolling Hough Transformation, also correlate with the magnetic field direction indicated by Planck dust polarization \citep{2014ApJ...789...82C,2015PhRvL.115x1302C}.

We do not know if such filamentary structures are the dominant contribution to the polarization foreground.  There is certainly evidence for filamentary structure in data \citep{2016A&A...586A.141P} as well as in simulations of the interstellar medium \citep{2005A&A...436..585D,2013A&A...556A.153H}, but there is no clear consensus on their origin or evolution \citep{2018arXiv181010014M,2019A&A...621A...5O}.  The purpose of this paper is to explore what polarization power spectra are possible for filaments, and what the observed power spectra can tell us about their physical properties.  


Other aspects of the dust polarization also need physical explanations.  
Both $E$-mode and $B$-mode spectra follow power laws ($C_\ell \propto \ell^\alpha$), with approximately the same slope,  $\alpha_{BB} = -2.42 \pm 0.02$ and  $\alpha_{EE} = -2.45 \pm 0.03$. \citep{2016A&A...586A.133P,2018arXiv180104945P}
 There is a positive correlation between dust intensity and $E$-mode
 polarization \citep[noted by][]{2017ApJ...839...91C}, with correlation
 coefficient $r_{TE}=0.357 \pm 0.003$ \citep{2018arXiv180104945P} and
 significant scatter depending on the sky area but little evidence for scale
 dependence.  Perhaps more intriguing is a parity-violating, positive $TB$
 correlation \citep{2018arXiv180104945P}.
  Finally, the amplitude of dust polarization power correlates to intensity in
  patches, roughly as $\langle I\rangle_{\rm patch}^{1.9}$, for both $E$ and
  $B$ \citep{2016A&A...586A.133P}.

 A few works have already tried to address these observations.
 \citet{2017ApJ...839...91C} examined the dust polarization power spectra of slow, fast, and Alfv\'en MHD waves in terms of two parameters: the ratio of gas to magnetic pressure, and the anisotropy of the MHD modes around the background field direction.
 They found two regions of parameter space that can account for the $E$ to $B$ ratio and positive $TE$ correlation but judged that these scenarios are unlikely due to the uniformity of the polarization power spectrum across the sky, and instead suggested that Planck may be seeing large scale displacements that are driving the turbulent ISM, rather than the turbulence itself.
 
On the other hand, \citet{2017MNRAS.472L..10K} argued with a similar analysis that the observed $E/B$ power ratio can be realized in an MHD model, so long at the turbulent flow is sub-Alfv\'enic. \citet{2018MNRAS.478..530K} extended this analysis to examine the $TE$ correlation and synchrotron emission.

Other works approach the problem using MHD simulations. For the most part, the ISM is filled with trans- and super-sonic flows, which are non-linear \citep[e.g.][]{2004ARA&A..42..211E,2010ApJ...708.1204B}. 
Both \citet{2018PhRvL.121b1104K} and  \citet{2019arXiv190107079K} made MHD simulations of  the ISM, and modeled the dust polarization
signals.  Both works find slopes and power ratios that are
reasonably close to the observed values, but the slopes
are especially sensitive to the masking procedure.
What MHD simulations do not provide is a straightforward and direct way to understand why these polarization properties arise.

Here we seek to gain physical intuition with very simple models of polarized filaments.  We compute their temperature and polarization power spectra using a method akin to the cosmological halo model \citep[e.g.][]{2000MNRAS.318..203S,2002PhR...372....1C}.  However, instead of spherical halos, we use magnetized, prolate-spheroidal filaments as the basic ingredients, and integrate over their population.

We organize this paper so that, in Section~\ref{sec:method}, we describe our formalism for characterizing the filament signal and for computing the power spectra.  In Section~\ref{sec:results}, we show the power spectra and discuss how the parameters of the filament population affect them.  In Section~\ref{sec:discussion}, we conclude  and discuss the implications and possible future directions.  
An appendix describes how the distributions of filament and magnetic field orientations in three dimensions appear when projected  onto the plane of the sky.

\section{Method}\label{sec:method}

We define a projected filament profile, $f(\bm{x})$, upon which we paint the temperature (i.e. intensity) and polarization signals.  Thus the temperature profile is:
\begin{equation}
  T(\bm{x}) = T_0 f(\bm{x}),
\end{equation}
for sky position $\bm{x}$.  
We model the polarization with an overall polarization fraction and polarization direction.  In terms of the Stokes parameters, the polarization for a filament is
\begin{eqnarray}
  X(\bm{x}) &=& (Q+iU)(\bm{x}) \\ \nonumber
  &=& f_{\rm pol} \exp(2i \psi_{\rm pol}) T_0 f(\bm{x}). 
\end{eqnarray}
In the HEALPix polarization convention \citep{2005ApJ...622..759G}, the $+x$-axis points south and the polarization angle $\psi_{\rm pol}$ increases east of south.  
Because we will integrate over angles in our computation of the power spectrum (and because $E$ and $B$ fields are coordinate independent) we can analyze a filament that has its long axis aligned (in projection) with the $x$-axis without loss of generality.
For simplicity, we assume that the intrinsic, microphysical contribution to $f_{\rm pol}$ is common to all filaments, although we will account for geometrical and projection effects in this work.
If the long axis of that filament were aligned with the local magnetic field, the precession of the dust grains would cause the polarization angle to be $\psi_{\rm pol} = 90^\circ$, perpendicular to the filament axis.

\begin{figure}
  \begin{center}
  \includegraphics[width=\columnwidth]{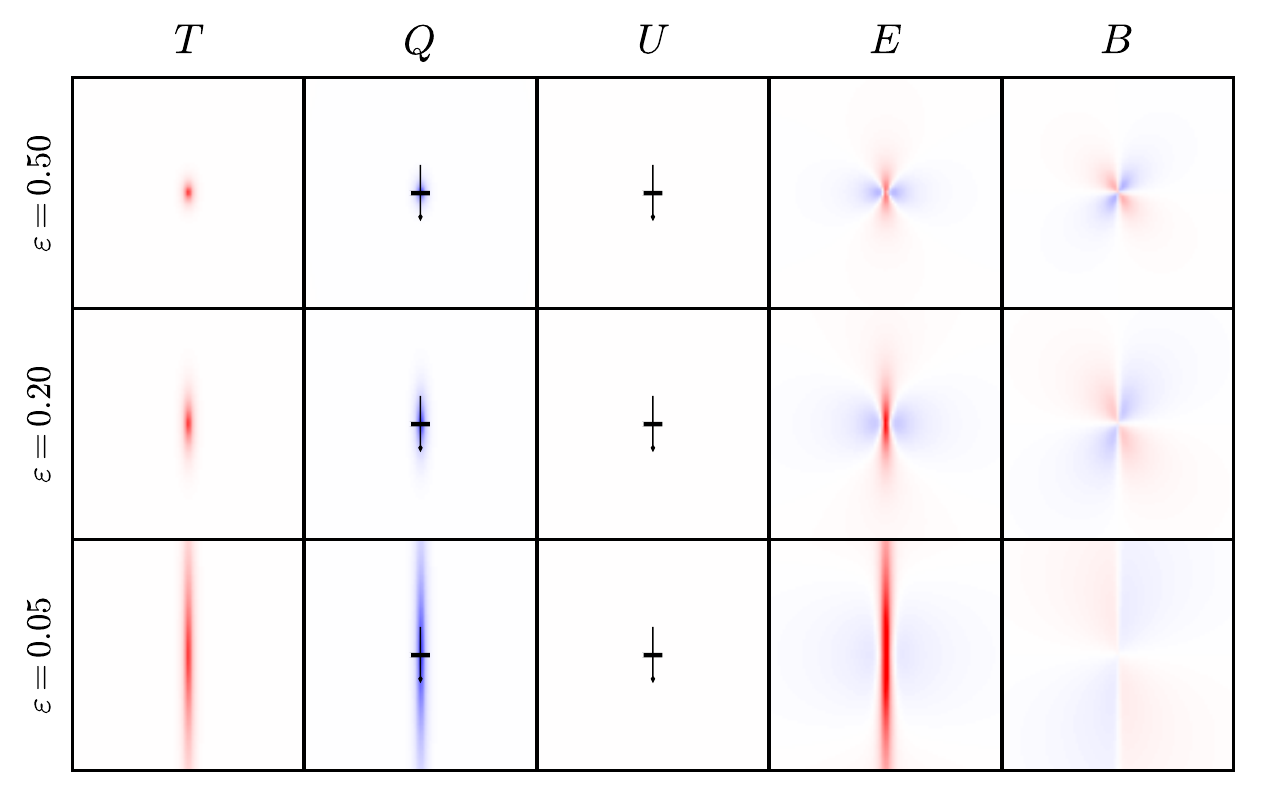}
  \end{center}
  \caption{Stokes parameters and scalar polarization quantities for idealized filaments as we alter the aspect ratio.  The magnetic field (thin arrow) is vertical, so the polarization direction is horizontal (thick line), making $Q<0$ and $U=0$.  In sky convention, north is to the top and east is to the left.\label{fig:filament}  The filaments are 2, 5, and 20 times longer than they are wide (axis ratio $\epsilon = 0.5, 0.2,0.05$).  Scalars $E$ and $B$ are on the same color scale, which has half the range of the $Q$ scale.  The $T$ scale differs from $Q$ by an arbitrary polarization fraction.}
\end{figure}

Working in the flat sky approximation, the Fourier components of the scalar polarization modes are:
\begin{equation}
  (E+iB)(\boldell) = \exp(-2i \phi_{\boldell}) X(\boldell).
\end{equation}
Fig.~\ref{fig:filament} shows the Stokes $T,Q,U$ and scalar $E,B$ quantities on the sky for sample, north--south filaments with $\psi_{\rm pol}=90^\circ$, so the magnetic field is parallel to the filament direction and the polarization is perpendicular. (Our choice of  coordinates implies that Stokes $U$ is zero in these cases.)   When the magnetic field aligns with the filament direction, \citet{2018arXiv180711940R} pointed out that the real-space kernels for the $E/B$ signals show immediately that the $E$-type polarization is positive along the filament, regardless of its orientation.  Since the temperature signal is also strong there, such filaments naturally yields a strong and positive $TE$ cross-correlation, as observed in the Planck data. The same work showed that the $B$ signal is concentrated at the ends of the filament, so filaments with long and thin aspect ratios will have less $B$ power relative to $E$ power than more squat ones.

\begin{figure}
\includegraphics[width=\columnwidth]{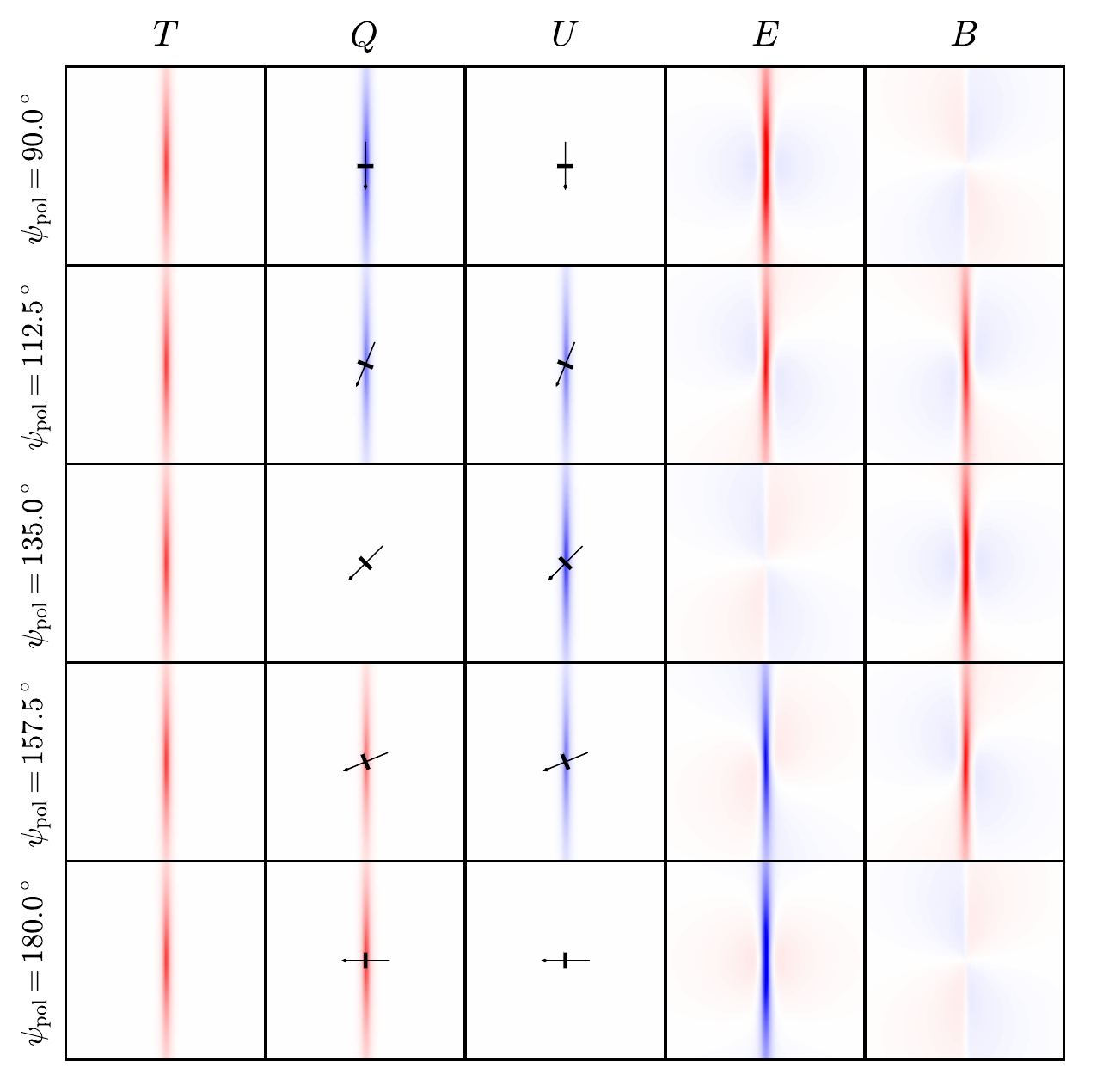}
\caption{Like Fig.~\ref{fig:filament}, but showing the polarization quantities as we alter the magnetic field direction (and hence the polarization angle), for a fixed filament orientation.  The $Q$ and $U$ Stokes parameters follow HEALPix polarization convention for a north--south filament, but the $E$ and $B$ fields are coordinate independent and appropriate for any orientation.  Note that filaments with aligned magnetic fields ($\psi_{\rm pol} = 90^\circ$) have zero $TB$ correlation.  Filaments with relative polarization angles  $90^\circ < \psi_{\rm pol} < 180^\circ$ have positive $TB$ correlations, as depicted, while those with $0^\circ < \psi_{\rm pol} < 90^\circ$ have negative $TB$ correlations (not shown).}\label{fig:pol_angle}
\end{figure}
By parity symmetry, the $TB$ and $EB$ cross-correlations are zero when the polarization is perpendicular to the filament (i.e.\ $\psi_{\rm pol}  =90^\circ$).  In Fig.~\ref{fig:pol_angle} we show how the $E$ and $B$ patterns transform into each other (and change sign) as $\psi_{\rm pol}$ varies away from $90^\circ$.  In these cases, the $TB$ and $EB$ correlation can be non-zero for individual filaments, but so long as the average $\langle \psi_{\rm pol} \rangle =90^\circ$, there will be no overall cross-correlation for the whole population.

\subsection{Projection on the sky}

\begin{figure}[t]
  \begin{center}
    \includegraphics[width=\columnwidth]{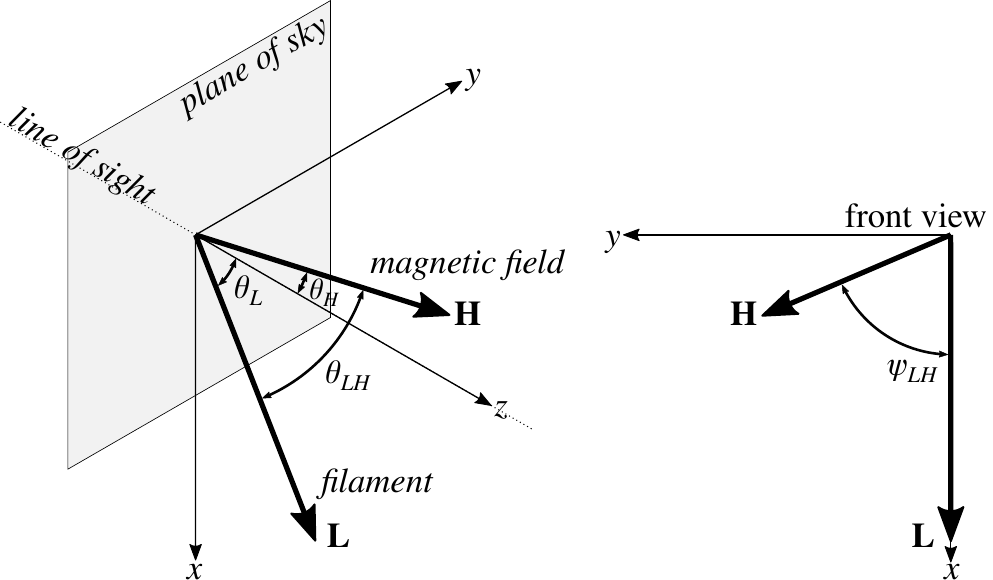}
  \end{center}
  \caption{Geometry of the filament direction and the magnetic field. The long axis of the filament $\mathbf{L}$ points in the $x$--$z$ plane at an angle $\theta_L$ from the line of sight. The magnetic field $\mathbf{H}$ has an angle from the line of sight of $\theta_H$. The angular misalignment between the field and the filament is $\theta_{LH}$.  The front view gives the projection of the misalignment onto the plane of the sky, $\psi_{LH}$.}
  \label{fig:anglefig}
\end{figure}

We next discuss the projection of a three-dimensional filament onto the plane of the sky.
Many important quantities depend on the angle to the line of sight of (1) the long axis of the filament ($\theta_L$) and (2) the magnetic field vector ($\theta_H$).\footnote{Elsewhere in the ISM literature, the angles are often given with reference to the plane of the sky, e.g. $\gamma_H = 90^\circ - \theta_H$ and so on.}  Another important quantity is the the plane-of-sky projection of the angle between these vectors ($\psi_{LH}$), which controls the polarization angle and the amounts of $E$/$B$ polarization present.  We depict these angles in Fig.~\ref{fig:anglefig}.  If the magnetic field direction aligns somewhat with the filament direction, as is the case in strong-field MHD, all these angles will be correlated.

We assume that on average, the filaments align with the local magnetic field.
In the appendix, we use simple geometry to compute the distribution of
the magnetic field projection angle $\theta_H$ and relative orientation angle $\psi_{LH}$ as a function of $\theta_L$.
  We base the distribution on the assumption of a Gaussian distribution for the angle ($\theta_{LH}$) between the filament and the magnetic field in three dimensions, characterized by the dispersion ${\rm RMS}(\theta_{LH})$.

  The field angle $\theta_H$ is correlated with $\psi_{LH}$, so our numerical procedure yields the tabulated joint distribution,
  \begin{equation}
    p(\psi_{LH},\theta_H | \theta_L).
  \end{equation}
  This distribution centers on aligned filaments ($\psi_{LH} = 0^\circ,\theta_H = \theta_L$), and the distribution for $\psi_{LH}$ broadens for filaments nearly along the line of sight.  Its precise form is not vital for this discussion and is plotted in the appendix in Fig.~\ref{fig:psiLH_thetaH_dist}.
  
On the other hand, the probability distribution for the line-of-sight angle of randomly oriented filaments is determined purely by geometry,
\begin{equation}
  p(\theta_L) = \sin \theta_L,
\end{equation}
for $\theta_L \in [0,180^\circ]$.

These quantities relate immediately to the polarization.  Although the dust polarization fraction depends on the microphysical details of the emission, it has a geometric dependence like  $f_{\rm pol} \propto \sin^2 \theta_H$ \citep{2000ApJ...544..830F}.  Meanwhile, the polarization angle for a filament projected along the $x$-axis is $\psi_{\rm pol} = \psi_{LH} + 90^\circ$.

\begin{figure}
  \includegraphics[width=\columnwidth]{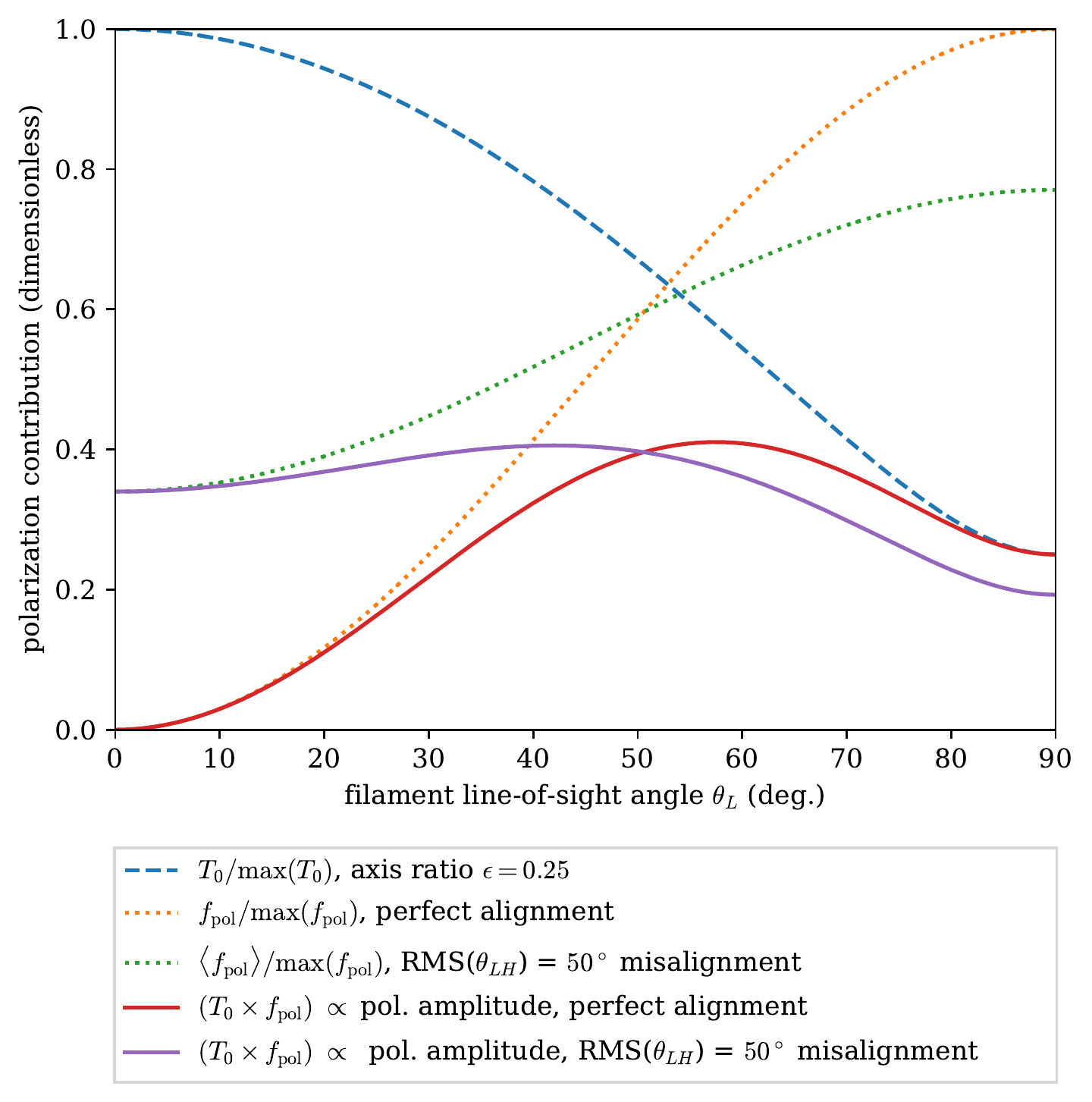}
  \caption{Total intensity, polarization fraction, and polarization amplitude dependence on the filament orientation.    Filament oriented along the line of sight have $\theta_L = 0^\circ$, while filaments in the plane of the sky have $\theta_L = 90^\circ$.   The filament is modeled with an axis ratio $\epsilon = 0.25$, and the filament direction ($\theta_L$) and the magnetic field direction ($\theta_H$) are either perfectly aligned or stochastically misaligned in three-dimensions. }\label{fig:pol_vs_thetaL}
\end{figure}

We model the filament as a prolate spheroid, and label the major axis as $L_a$ and the minor axis as $L_b$.  The axis ratio is thus $\epsilon = L_b/L_a < 1$.  The column density (and therefore the surface brightness and ultimately the observed temperature perturbation) is proportional to the density and the line of sight distance through the filament, and so (approximately)
\begin{eqnarray}
  T_0 &\propto&  \rho_0 \left( L_a^2 \cos^2 \theta_L + L_b^2 \sin^2 \theta_L \right)^{1/2} \\ \nonumber
  &\propto&  \rho_0 L_a  \left( \cos^2 \theta_L + \epsilon^2 \sin^2 \theta_L \right)^{1/2}.
  \label{eq:column_density}
\end{eqnarray}
where $\rho_0$ is a characteristic density for the filament.
So all else being equal, a filament that lies along the line of sight will have the greatest column density and the brightest temperature signal.  On the other hand, $f_{\rm pol} \propto \sin^2 \theta_H$, so if the magnetic field lies along the line-of-sight, there is no polarization.  The polarization fraction is maximum when the magnetic field is perpendicular to the line of sight.

Fig.~\ref{fig:pol_vs_thetaL} relates the column density and polarization fraction effects of the line-of-sight angle.  It also shows that since the polarized amplitude depends on the product of these two, the filaments with the brightest polarization are inclined, but not perpendicular, to the line of sight.  The polarization maximum depends on axis ratio through its impact on the column density.  Nearly round filaments have the polarization maximum when oriented near  90$^\circ$ to the line of sight, while in the limit of thin filaments ($\epsilon \rightarrow 0$) the polarization maximum orientation approaches $\theta_L = 45^\circ$ for perfect magnetic field alignment.  If there is significant misalignment of the magnetic field and filament directions, the situation can become more complicated, depending on the particular combination of axis ratio and misalignment dispersion.  In such cases, filaments along the line of sight can have significant polarization.  Still, the typical line-of-sight orientation angle for maximum polarization, averaging over the magnetic field directions, is around $\theta_L = 45^\circ$.

We compute the filament's projected angular sizes along its two axes as if it were a cylinder.  These depend  its distance $R$ and are:
\begin{eqnarray}
  \Theta_a &=& \left( L_a^2 \sin^2 \theta_L + L_b^2 \cos^2 \theta_L \right)^{1/2} / R \\ \nonumber
  &=&  \left( \sin^2 \theta_L + \epsilon^2 \cos^2 \theta_L \right)^{1/2} L_a / R \\ \nonumber
  \Theta_b &=& L_b/R = \epsilon L_a/R.
\end{eqnarray}
Thus the projected axis ratio is
\begin{equation}
  \epsilon_\Theta = \Theta_b/\Theta_a = \frac{\epsilon}{\left( \sin^2 \theta_L + \epsilon^2 \cos^2 \theta_L \right)^{1/2}}
\end{equation}
which goes to unity for filaments along the line of sight, and to the true value ($\epsilon$) for filaments perpendicular to the line of sight.

\subsection{Filaments in Fourier space}

For several terms in our power spectrum calculation, we need the Fourier transform of the projected filament profile:
\begin{equation}
  f(\boldell) = \int d^2x\  f(\bm{x}) \exp(-i \boldell \cdot \bm{x}).
\end{equation}
Rather than project rays through a 3-dimensional model to obtain the filament profile, we make a simplifying assumption for computational efficiency.  From the size and orientation of a filament, we take the angular dimensions and compute under the assumption that the profile is a distortion from an axisymmetric function $g$:
\begin{eqnarray}
  f(x,y) =& g(x/\Theta_a , y/\Theta_b)\ & \\ \nonumber =& g(x^*, y^*) \quad & = \ g(r)
\end{eqnarray}
where $\Theta_a,\Theta_b$ are the semi-major and semi-minor axis of the elliptical distortion.  As we stated, by convention and without loss of generality, we orient the long axis of the filament along the $x$-axis.

Then the transform of $f$ is simply related to the transform of $g$:
\begin{eqnarray}
  f(\boldell) &=& \Theta_a \Theta_b \int d^2x^*\  g(\bm{x}^*) \exp(-i (\Theta_a\ell_x x^* + \Theta_b \ell_y y^*)) \nonumber \\
  &=& \Theta_a \Theta_b\,  g( \ell^* )
\end{eqnarray}
where
$  \ell^*(\boldell) = (\Theta_a^2 \ell_x^2 + \Theta_b^2\ell_y^2)^{1/2}$
and
\begin{eqnarray}
  g(\ell^*) = \int d^2x^* \  g(\bm{x}^*) \exp(i \boldell^* \cdot \bm{x}^*) \\ \nonumber
  = 2\pi \int dr \ r\  g(r) J_0(\ell^* r).
\end{eqnarray}
The input profile $g(\mathbf{x}^*)$ is real and even, and so the Fourier transform is too.  The power spectra we find are not very sensitive to the profile that we use.  In this work we have used an exponential for the basic filament profile ($g(r) = \exp(-r)$), but we have checked our best-fitting power spectrum model with a Gaussian profile ($g(r) = \exp(-r^2/2)$) and a Plummer profile ($g(r) = (1+r^2)^{-5/2}$), and find the same results. 







\subsection{Parameters and one-filament term}

For a parameterized set of filament properties, $$\alpha = ( L_a, L_b, \psi_{LH}, \theta_L, \theta_H,R,\dots),$$ we can write the number density distribution $n(\alpha)$, so that the average number of filaments in a realization of the sky is
\begin{equation}
  \langle N \rangle = \int d\Omega\  d\alpha\ n(\alpha)
\end{equation}
where the integral is over
\begin{equation} d\alpha =  dL_a dL_b d\psi_{LH} d\theta_H d\theta_L dR. \end{equation}

Expressed another way, $n(\alpha) =  \langle N \rangle p(\alpha)$, where the normalized probability distribution of the filament population is
\begin{equation} p(\alpha) = p(L_a,L_b)p(\psi_{LH},\theta_H | \theta_L) p(\theta_L)p(R)  \end{equation}

This integral over the population is at least  six dimensional.  For a screen at a distance $R$, it is five dimensional integral.  Since the angular power spectrum for foregrounds is a power law, if we can reproduce it on a single screen, putting that screen at different distances will maintain the same power spectrum.  If we further fix the physical aspect ratio of the filaments, it is a four dimensional integral, over $L_a, \psi_{LH}, \theta_L, \theta_H$.  (The projected aspect ratio will still vary with the line-of-sight angle $\theta_L$.)

The power spectrum contributions from filaments correlated with themselves are:
\begin{eqnarray}
C_\ell^{TT} &=& \frac{1}{2\pi} \int d\phi_\ell \int d\alpha\  n(\alpha) \  | T(\boldell,\alpha)|^2, \\ \nonumber
C_\ell^{EE} &=& \frac{1}{2\pi} \int d\phi_\ell \int d\alpha\  n(\alpha) \  | E(\boldell,\alpha)|^2, \\ \nonumber
C_\ell^{BB} &=& \frac{1}{2\pi} \int d\phi_\ell \int d\alpha\  n(\alpha) \  | B(\boldell,\alpha)|^2, \\ \nonumber
C_\ell^{TE} &=& \frac{1}{2\pi} \int d\phi_\ell \int d\alpha\  n(\alpha) \  T(\boldell,\alpha) E(\boldell,\alpha)^*.
\end{eqnarray}
Similar expressions hold for the other cross correlations, but these vanish if the orientations of the filaments are random.  These power spectra computations are directly analogous to the 1-halo term in the cosmological halo model \citep{2000MNRAS.318..203S}.

%
%

\section{Results}\label{sec:results} 

There are clear relationships between the physical properties of the filaments and the temperature and polarization power spectra that they produce.  The slopes of the power spectra are determined primarily by the size distribution of filaments, with other effects responsible for the smaller differences between the components. The ratio of $BB/EE$ power is determined mostly by the aspect ratio of the filaments and somewhat by the misalignment of the filament directions to the background magnetic field.  These same factors also determine the cross correlation $r^{TE}_\ell$,
but here misalignment is much more important.  They also affect the $TE/EE$ power ratio, but this quantity is more directly affected by the overall polarization fraction.

\subsection{Power spectrum shape}

We can relate the slope of a power law spectrum to scaling relations for parameters in the filament profiles.  This allows us to place constraints on the distribution of filament sizes and the scaling of other parameters.  
For a generic parameter $\alpha_0$, if the filament's contribution to the power spectrum scales as
\begin{equation}
  C_\ell \propto \int d\alpha_0\  n(\alpha_0)\times \alpha_0^q \,F( \alpha_0^r\,  \ell) \label{eqn:shape_scaling}
\end{equation}
for any function $F$, and furthermore if the weighting distribution for the parameter is a power law, $n(\alpha_0) \propto \alpha_0^p$, then we can rescale the integration with a straightforward substitution, $u=\alpha_0 \ell^{1/r}$:
\begin{eqnarray}
  C_\ell &\propto& \ell^{-(p+q+1)/r} \times \\ && \nonumber \quad \int d(\alpha_0 \ell^{1/r})\   (\alpha_0 \ell^{1/r})^p \times (\alpha_0 \ell^{1/r})^q \,F( ( \alpha_0\,  \ell^{1/r})^r) \\ \nonumber
  &\propto& \ell^{-(p+q+1)/r} \times  \int du\   u^p  u^q \,F( u^r).
\end{eqnarray}
The integral no longer has any multipole dependence and evaluates to some constant value, whatever the details of $F$.  Thus we are left with a powerlaw power spectrum with
\begin{equation}
  C_\ell \propto \ell^{-(p+q+1)/r}
\end{equation}
This argument holds not just for filaments, but for any signal with a power-law power spectrum that is built from a set of objects that are similarly related to each other, so long as they are weighted by powerlaw scalings and distributions.  So if we observe a powerlaw spectrum with $C_\ell \propto \ell^s$, it implies that the parameter distribution's index is $p = -rs - q - 1$, regardless of the objects' profiles.

We walk through this scaling argument for a simple (and unrealistic) case---with plane-of-sky filaments with identical surface brightnesses ($T_0$ is the same for all filaments) and a constant projected-axis-ratio ($\Theta_b = \epsilon \Theta_a$)---and analyze the distribution for $\Theta_a$, the angular size of filaments.  For the filament Fourier transform, we  have $f(\boldell) \propto \Theta_a^2 g(\boldell^*)$ with $\boldell^* \propto \Theta_a$.  The power spectrum contribution is proportional to $f^2$, so comparing the scaling for angular size parameter $\Theta_a$ to equation~\ref{eqn:shape_scaling}, we find $q=4$ and $r=1$.
  In the polarization case, to reproduce $C_\ell \propto \ell^{-2.4}$ (meaning $s=-2.4$), the number density distribution of such objects on the sky must approximately scale like $n(\Theta_a) \propto \Theta_a^p$ where $p = 2.4 - 4 - 1 = -2.6$.  Indeed this yields the desired power spectrum slope when calculated in our model.

\begin{figure}
  \begin{center}
    \includegraphics[width=\columnwidth]{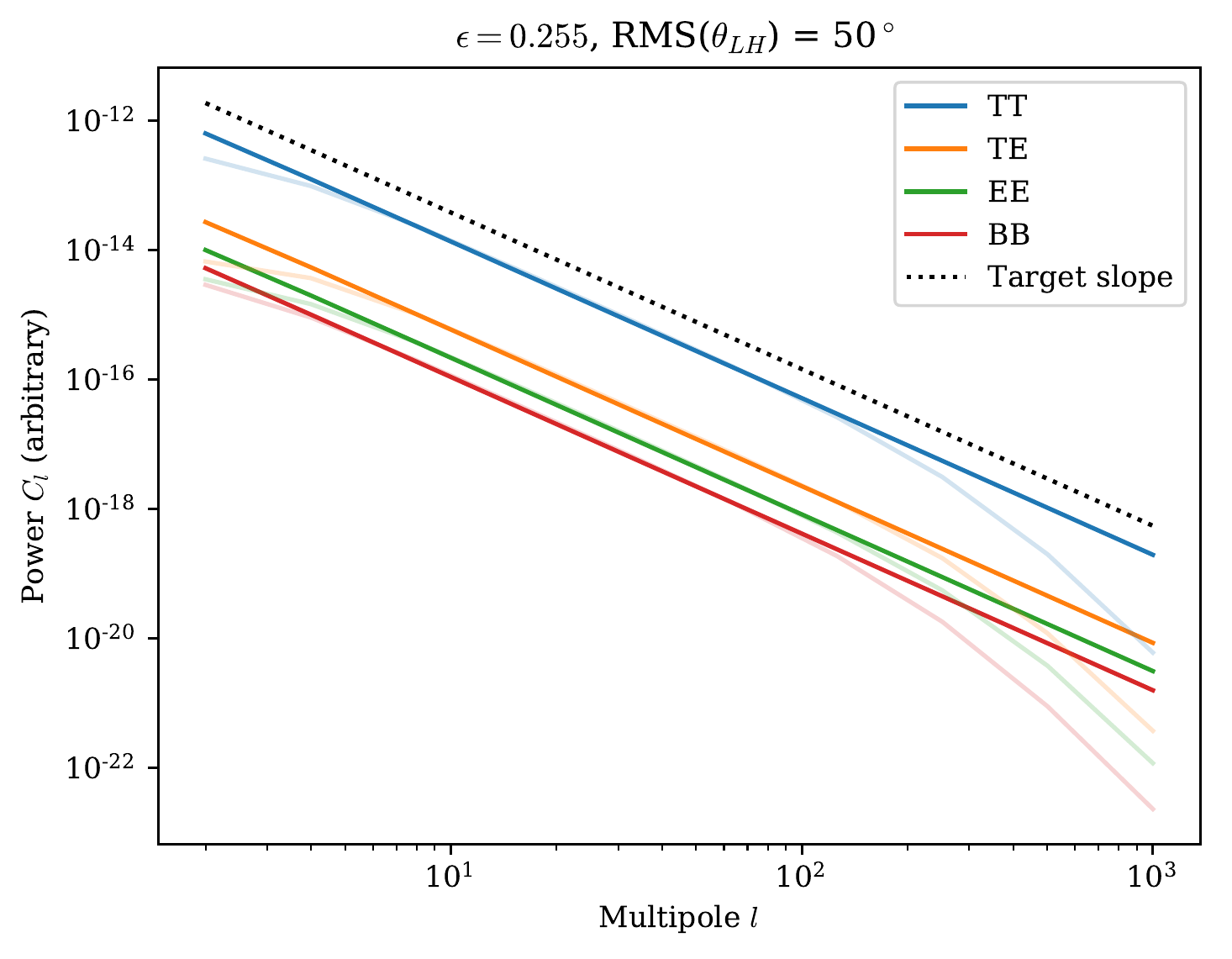}
  \end{center}
  \caption{Power spectra slopes of temperature and polarization are set chiefly by the distribution of filament lengths, which has been chosen here so that the slopes match the dotted line, $C_\ell \propto \ell^{-2.4}$.  The ratios between the power spectra are set by the overall polarization fraction, the aspect ratio of filaments, and the misalignment between the filaments and their local magnetic field.  When maximum and minimum sizes truncate the distribution of the filaments, the power falls below the targeted slope (semi-transparent colors).}
  \label{fig:slope}
\end{figure}
In a more realistic case, with three-dimensional filaments, we can make a similar argument to deduce the distribution of filament lengths.  Surface brightness depends on column density, which is proportional to length (after integrating out any distribution of axis ratios---compare equation~\ref{eq:column_density}---and assuming a density normalization independent of length).  The solid angle scales like length squared.  After squaring those three powers during the computation of the power spectrum, the overall scaling is $q=6$.  The multipole $\ell$ scaling should also go like length, so $r = 1$, the same as the plane-of-sky case above.  So with no other dependence on length, we should have distribution of lengths  $n(L_a) \propto (L_a)^p$ where $p = 2.4 - 6 - 1 = -4.6$.\footnote{If the column density normalization depends on length, this procedure yields a net distribution that is a product of the size distribution and the density squared distribution (both as a function of length).  Such a case could arise, for example, if small filaments are collapsed versions of large ones and have higher density.}

We have verified that this distribution produces the proper slope  in Fig.~\ref{fig:slope}.   All the temperature and polarization spectra have the specified slope in common.  The complications of the modeling of the three dimensional orientation are not important to the slope, only the weighting and distribution of filament size.

Other than the slope, there are not clear features in the Planck-measured spectra.  We note that features in the distribution of filament sizes would break the powerlaw behavior of the resulting spectra.  For example, if we impose a maximum filament size (semi-transparent lines in Fig.~\ref{fig:slope}), it causes the low-$\ell$ behavior of $C_\ell$ to deviate: at scales much larger than the filament, the temperature spectrum adopts the flat, white, Poisson spectrum of point sources.  The $EE$ and $BB$ spectra flatten the same way, and on scales large compared to the filaments, the aspect ratios of the filaments become unimportant and the amount of power in $EE$ and $BB$ equalize.  The $TE$ cross-correlation falls off at large scales, possibly because the positive and negative contributions to $E$ are being averaged over.  If, on the other hand, we impose a minimum filament size, it causes the high-$\ell$ behavior to deviate: the spectrum will drop off  with increasing $\ell$, where there is no more contribution to the power.


In yet more realistic cases, we can make the slopes of  all the temperature and polarization spectra differ.  For example, this can happen if there is another effect that changes the size scaling of the polarization relative to the temperature.  For example, to make polarization slope shallower than the temperature slope, one could make smaller objects more polarized than large objects, or if the aspect ratio of filaments changes as a function of size.
 
Considering the Planck data, it is not immediately clear what conclusion to draw.  \citet{2018arXiv180104945P} quote $EE, BB, TE$ slopes for the dust foreground, but not the $TT$ slope.  Using our own tools, we have computed the $TT$ power spectrum based on the Planck data, and find a $TT$ slope that is about $-2.6$, somewhat steeper than the polarization spectra at $-2.4$ to $-2.5$.  Like \citet{2018arXiv180104945P}, for the mask we used the LR71 polarization mask supplemented with a point source mask (based on intensity maps), resulting in a mask with effective $f_{\rm sky} \simeq 0.6$.
We can approximately reproduce a $-2.6$ temperature slope and $-2.4$ polarization slope with $n(L_a) \propto L_a^{-4.4}$ and $f_{\rm pol} \propto L_a^{-0.08}$.  This argues that in the unmasked region, the polarization is higher in smaller filaments.

However, this conclusion may not be correct because it depends sensitively on the mask.  We reasoned that small filaments, oriented along the line of sight, might look like a point source and be included in the masked area.  These end-on filaments could also have low polarization (note Fig.~\ref{fig:pol_vs_thetaL}), and excluding them might not have much effect on the polarization results.  Thus for comparison, we recomputed the $TT$ spectrum with different masks.   When we use only the polarization LR71 mask without removing the additional point sources from the intensity map, we get a shallower $TT$ spectrum with slope $-2.5$. When we use a mask that keeps the same large scale features but does not mask any point sources (Planck's publicly available GAL70 mask) we find a $TT$ slope of $-2.1$, notably shallower than the polarization spectra.  The polarization spectra change somewhat between these masks, but the changes in the polarization slopes are small compared to the change in the TT slopes.
Some of the masked sources are extragalactic, so this slope with all point sources unmasked is probably too shallow to describe the ISM component, but can serve as a bound.  The upshot is that we are not certain whether the spectrum for all filaments is steeper or shallower in  temperature than polarization, and so it is difficult to draw conclusions on the size dependence of the polarization fraction.

Another feature of the Planck data is the differing slopes in $E$ and $B$.    We can reproduce this feature by varying the  aspect ratio as a function of filament size.   For the LR71 mask in \citet{2018arXiv180104945P}, the slopes for $(BB, EE, TE)$ are roughly $(-2.5, -2.4, -2.5)$ and we found a $TT$ slope of $-2.6$.  So $BB$ is steeper than $EE$, which should happen if smaller filaments are proportionally thinner than longer ones.  Modifying the aspect ratio in this way also affects the $TT$ slope, breaking the simple relation that we saw earlier in this section.  By trial and error, we found that this set of slopes are approximately reproduced with the following parameter dependence: $\epsilon \propto L_a^{0.1}$, $n(L_a) \propto L_a^{-4.45}$, and $f_{\rm pol} \propto L_a^{-0.1}$. Here we are simply exploring what is possible, but the relationship between the measured slopes and these filament parameters should be made more systematic and quantitative.

In light of these complications and uncertainties, in what follows we keep a common slope of $-2.4$ for all the temperature and polarization components while we explore their other parameter dependences.

\subsection{BB/EE power ratio}

\begin{figure}
  \begin{center}
     \includegraphics[width=\columnwidth]{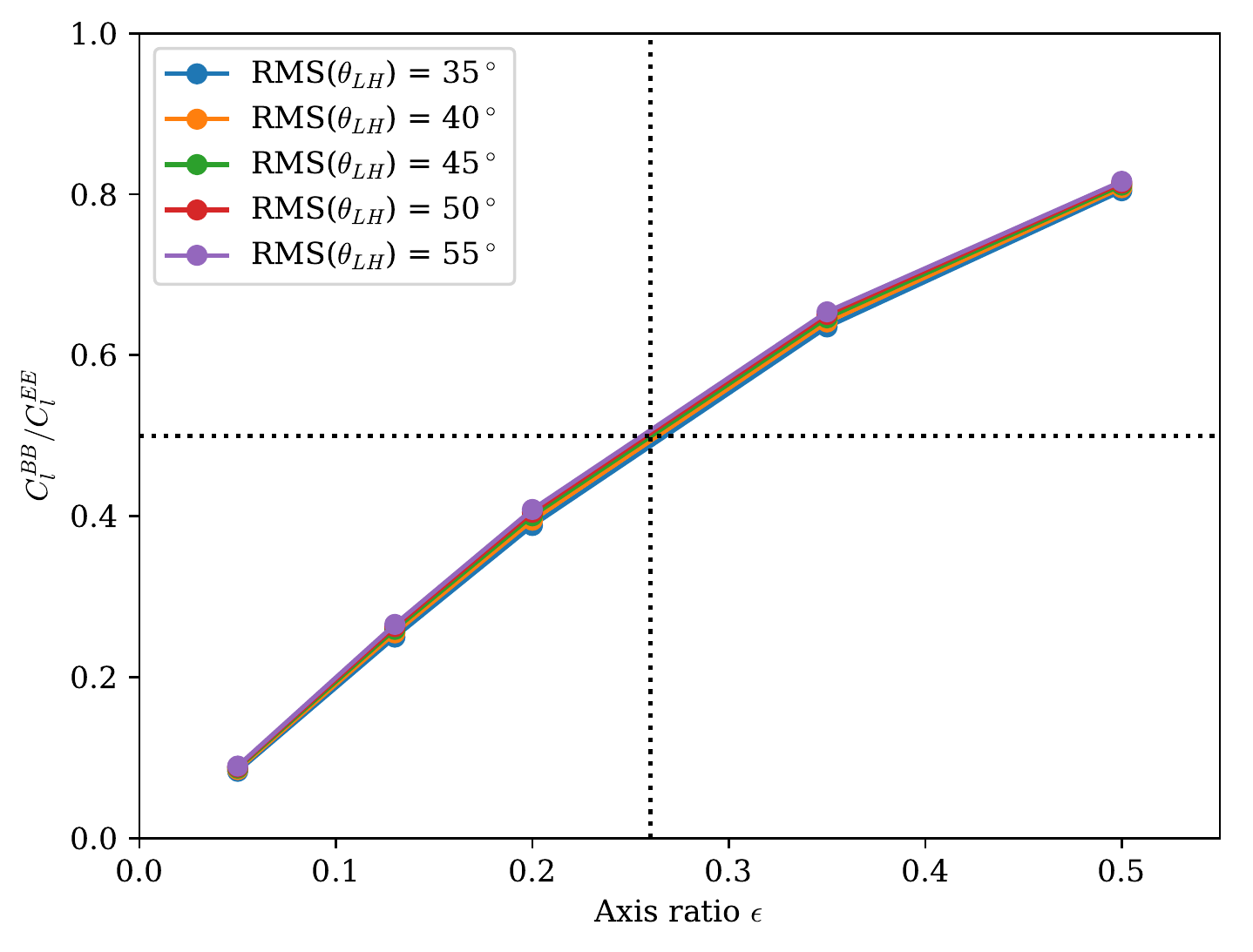}    
  \end{center}
  \caption{Ratios of $BB$ to $EE$ power as a function of the physical filament axis ratio.   Long and thin filaments ($\epsilon$ small) have less $B$-mode power than $E$-mode power. Short and squat filaments (aspect ratio $\epsilon$ close to unity) have $B$ power close to the $E$ power.  An aspect ratio of about $\epsilon = 0.26$ can reproduce the Planck-observed ratio of about one half, but this can be traded off against a slight dependence with the dispersion in the misalignment angle $\theta_{LH}$ between the filament direction and the magnetic field direction in three dimensions.}
\label{fig:EEBBrat}
\end{figure}

The aspect ratio of the filaments is the major factor determining the ratio of $B$-mode power to $E$-mode power.  In Fig.~\ref{fig:EEBBrat}, we plot the power ratio against the aspect ratio for varying degrees of filament--magnetic field misalignment.  To reproduce the Planck-observed ratio of $\sim 0.5$, filaments need to have an aspect ratio $\epsilon$ slightly less than 0.26, so filaments must be slightly less than four times longer than they are wide.  If the model deviates too much from this ratio, the required magnetic field misalignment is made so large that the model has trouble fitting the $TE$ correlation.

It is difficult to compare this result quantitatively to the aspect ratios of observed filaments from the literature without making a detailed accounting of the filament selection function.  Projection effects will also tend to lower observed aspect ratios.  The stacked filaments in Fig.~7 of \citet{2016A&A...586A.141P} appear to have axis ratios not so far from what we are finding here.  The filaments identified by the Rolling Hough transformation  in \citet{2014ApJ...789...82C} on HI maps tend to be longer and thinner than this.

\begin{figure}
  \begin{center}
    \includegraphics[width=\columnwidth]{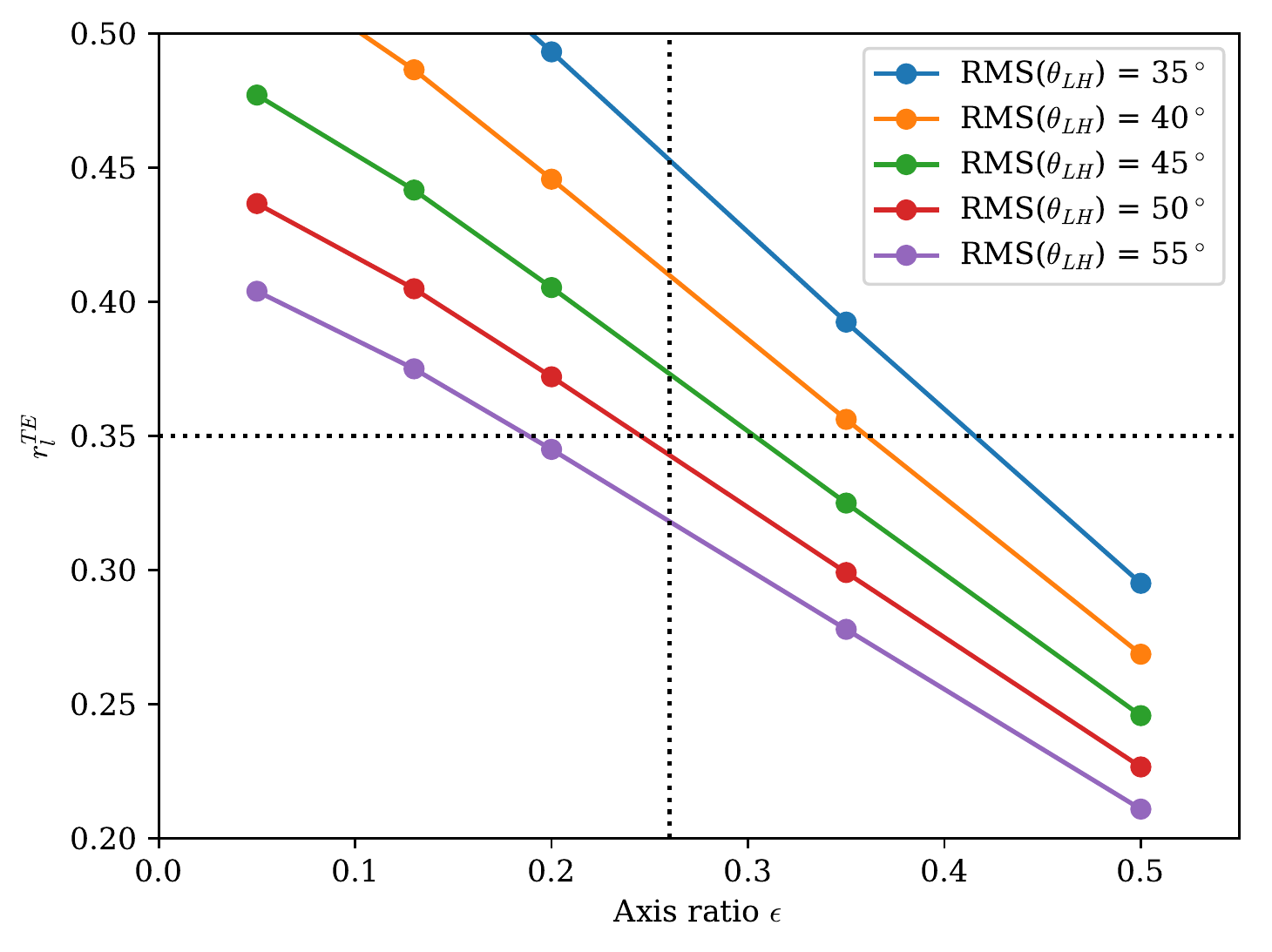}
  \end{center}
  \caption{Correlation coefficient of $TE$ power as a function of the axis ratio. Thinner filaments (small $\epsilon$) have stronger $TE$ correlations for various magnetic field--filament misalignments.   Since the $B$-to-$E$ power ratio requires $\epsilon \approx 0.26$ (Figure \ref{fig:EEBBrat}), the $TE$ correlation is diagnostic of the misalignment necessary to produce the Planck-observed $r^{TE} \approx 0.35$, which needs ${\rm RMS}(\theta_{LH}) \approx 50^\circ$.}\label{fig:rTE}
\end{figure}

\subsection{TE cross-correlation}

In the context of the filament model, we find that the level of the $TE$ cross correlation implies that filaments cannot be precisely aligned to their local magnetic field direction.  The correlation coefficient is defined as
\begin{equation}
  r_\ell^{TE} =   C_\ell^{TE}/\sqrt{C_\ell^{EE} C_\ell^{TT}},
\end{equation}
and perfect alignment of  the filaments and the fields causes far too much $TE$ correlation compared to the Planck observations.

The Planck dust data show $r_\ell^{TE} \approx 0.35$ with little scale dependence \citep{2018arXiv180104945P}.  Fig.~\ref{fig:rTE} shows that to match this, the field misalignment angle $\theta_{LH}$ must have an RMS dispersion of nearly $50^\circ$, while maintaining the axis ratio $\epsilon \approx 0.26$ needed to reproduce the $BB/EE$ power ratio.  If the misalignment dispersion is independent of filament size, as in our modeling, it causes no scale dependence: $r^{TE}_\ell$ is constant.

{Projection effects cause the distribution of the projected angle $\psi_{LH}$ to have a positive kurtosis (see appendix, Fig.~\ref{fig:psiLH_marg}), and so we can describe its dispersion in a few ways.  For the ${\rm RMS}(\theta_{LH}) = 50^\circ$ case, 68 percent of the probability is bounded by  $|\psi_{LH}| < 45^\circ$.  Alternatively, $[{\rm Var}(\psi_{LH})]^{1/2} = 48^\circ$.}  For comparison, \citet{2016A&A...586A.141P} fit a Gaussian a with $19^\circ$  dispersion (1$\sigma$)  to the projected field--projected filament histogram of relative orientations for the filaments they found.  Again this comparison is not direct because of selection effects.  Their Hessian-based selection of filaments would disfavor filaments with small projected aspect ratios (close to the line of sight), and such filaments can have the largest differences in the projected orientation.

The overall level of $C_\ell^{TE}$ (and the polarization spectra) depends on the polarization fraction.    The relative power ratio has a dependence like
\begin{equation}
  C_\ell^{TE}/C_\ell^{EE} \propto \langle f_{\rm pol} \rangle/\langle f_{\rm pol}^2 \rangle,
\end{equation}
while for the cross correlation it is
\begin{equation}
  r_\ell^{TE}  \propto \langle f_{\rm pol} \rangle/\langle f_{\rm pol}^2 \rangle^{1/2}.
\end{equation}
Thus a purely multiplicative rescaling of the polarization fraction affects the ratios of the power in $TT/TE/EE$ but not the correlation coefficient $r^{TE}$. 

To reproduce the Planck-measure ratio $C_\ell^{TE}/C_\ell^{EE} \sim 2.7$ (in our case that already fits $C_\ell^{BB}/C_\ell^{EE}$ and $r_\ell^{TE}$) requires $f_{\rm pol} = 0.15\, \sin^2 \theta_H$.
We have only modeled the polarization fraction amplitude and the geometric dependence on the magnetic field orientation, but in addition, the polarization fraction depends on grain geometry and small-scale turbulence \citep{2000ApJ...544..830F}, and filaments need not in reality have all the same intrinsic polarization fraction.

Our other tests have shown that the $TE$ correlations differ in their sensitive to intrinsic dispersion in the polarization fraction.  For example, the correlation $r^{TE}$ is not very sensitive to the maximum polarization fraction, but the power ratio is very sensitive to it: decreasing the maximum polarization fraction decreases $C_\ell^{TE}$ but decreases the denominator $C_\ell^{EE}$ more, and so raises the ratio.

\subsection{Parity violation: TB and EB}

One surprising finding in the \citet{2018arXiv180104945P} dust spectra is a non-zero $TB$ correlation, with $r_\ell^{TB} \approx 0.05$.  Like $r^{TE}$, the observed $r^{TB}$ correlation has little scale dependence (up to multipoles of several hundred).

Because non-zero $TB$ and $EB$ are parity-violating correlations, our model cannot reproduce them for randomly-oriented filaments.  To get a positive $TB$ correlation we would need to favor polarization angles in the range $\psi_{\rm pol} \in [90^\circ,180^\circ]$ relative to the filament direction (compare Fig.~\ref{fig:pol_angle}).  Equivalently, this corresponds to projected field angles in the range $\psi_{LH} \in [0^\circ,90^\circ]$.  Such an effect may be due to some large scale feature in the Galaxy's magnetic field or differential gas flow \citep[e.g.][]{2018arXiv180104945P,2019A&A...621A..97B}.

We can determine how far away from random this correlation is by artificially weighting the distribution of the projected misalignment angles, favoring the $\psi_{LH} > 0$ portion of the distribution of $p(\psi_{LH},\theta_H | \theta_L)$ over the $\psi_{LH} < 0$ portion, while keeping the same functional form.  We find that we can approximately reproduce the Planck measured $TB$ correlation by giving the preferred $\psi_{LH}$ directions about 55 percent of the total weight, rather than the 50 percent than comes naturally from randomly oriented filaments.  Similar to the $r^{TE}$ correlation (also set by field--filament misalignment), this effect is not scale dependent, and so $r^{TB}$ is constant to high $\ell$ in this model.

Our modeling comes from the 1-halo term only, and shows that the $TB$ correlation can be explained if filaments orientations in projection are slightly twisted counterclockwise from the projected local magnetic field.  Our model does not address the structure of that underlying field, but we may speculate that some differential, shearing hydrodynamic forcing could preferentially twist the filaments, according to our point of view, from the global mean field direction of the Milky way. 

\citet{2019A&A...621A..97B}, argue that the observed $TE$ and $TB$ correlations may be features of the large scale structure of the Galactic magnetic field.  They show that a helical component can create such correlations, but in their modeling, the correlation show a strong scale dependence, with $r^{TE,TB}_\ell$ falling substantially already by $\ell = 22$.  The Planck spectra have much flatter $r^{TE,TB}_\ell$ correlations, consistent with the filament modeling here.

Planck did not detect an $EB$ correlation, but since $E$ and $B$ both have a factor of the polarization fraction, we would  naturally expect this correlation to be smaller.  It may be there, hidden in the noise.   In the presence of a positive $TB$ correlation, in the context of the filament modeling, we would expect a positive $EB$ correlation too (including at high-$\ell$), and it should be a target for future experiments.  Both $TB$ and $EB$ dust correlations can potentially interfere with sky-calibration of the polarization angles of CMB-instruments \citep{2016MNRAS.457.1796A} or with CMB lensing reconstruction \citep[e.g.][]{2012JCAP...12..017F,2018JCAP...04..018C}.

\section{Conclusions}\label{sec:discussion}

We do not know how much of the dusty microwave polarization foreground is due to filamentary structure, but if it is a substantial portion, we can discern details of the filament population from the foreground power spectra.  We showed that the slopes of the power spectra relate to the distribution of lengths.  We showed that the $BB$/$EE$ power ratio relates to the filament axis ratio.  We showed that the $TE$ cross-correlation relates to the axis ratio and the RMS misalignment of filaments to the magnetic field.  We showed that $TB$ correlations could be caused by a slight preference for one handedness in the misalignment between the magnetic field and the filament orientation.

Despite its relative success in reproducing the features of the dust polarization power spectrum, this formalism lacks some essential features for modeling the real sky.  Foremost, this formalism includes only the one-halo term in the power spectra.  This is obviously an approximation, for the Planck data have shown that the Galaxy's projected magnetic field has coherent, large-scale features, and the HI-identified filaments are clearly correlated with it and with starlight polarization measurements \citep{2015PhRvL.115x1302C}.
On the other hand, the transition from one-halo-dominated to two-halo-dominated scales often leaves a mark on the power spectrum.  Since in the dust polarization spectra there are not clear features, like a break in the slope, we may speculate that the two-halo component may not be necessary to describe the main properties of the power spectra.
Inclusion of a proper two-halo formalism is complicated by the correlated direction dependence of the filaments.  We may be able to import some of the techniques developed to describe galaxy intrinsic alignments \citep[e.g][]{2010MNRAS.402.2127S}, since the mathematical description of the problem is similar.

We have not tried to systematically probe the parameter degeneracies or place proper uncertainties on any of the parameters of this filament model.  We can do this straightforwardly by interfacing the model with a Monte Carlo Markov Chain and developing a likelihood based on the Planck dust spectra. We plan to pursue this in further work.

By looking at observations and simulations of the filamentary ISM, we could attempt to verify some of the population statistics for filaments.  For example, we could compare the length distribution of actual or simulated filaments to that implied by the slope of the power spectra.

Because this filament model is non-Gaussian, we may be able to use it to design novel diagnostics to probe for residual foregrounds in surveys that aim for the primordial $B$-modes  \citep[in the spirit of][]{2014PhRvL.113s1303K,2016JCAP...09..034R,2018MNRAS.479.5577P,2019arXiv190104515C}.  Similarly, we could use this model to compute the four-point contributions to polarized CMB lensing estimators.  This could help place constraints on potential foreground contamination.  Such statistics may be sensitive to the internal density structure of the filaments is a way that the power spectrum is not.
 
Due to its flexibility, its ability to model the Planck dust polarization data, its ease of computation, and its straightforward interpretation, this filament model may become a useful tool in the study of CMB polarization foregrounds.

\section*{Acknowledgments}
KMH, AR, and DCC acknowledge support from the NASA ATP program under grant NNX17AF87G. KMH acknowledges support from the NSF AAG program under grant 1815887. AR acknowledges support from the European Research Council (ERC) under the European Union's Horizon 2020 research and innovation programme (grant agreement No 725456, CMBSPEC). We thank J.~Colin Hill, Susan Clark, Marc Kamionkowski, and Carlos Hervias-Caimapo for useful conversations.  We thank Francois Boulanger for providing access to masks used in the  \citet{2018arXiv180104945P} analysis.

\bibliography{ref}

\appendix

\section{Distributions of angles}

We describe the directions of (the long-axis of) the filament and the magnetic field with the coordinates in Fig.~\ref{fig:anglefig}.  The line of sight lies along the $z$-axis, while $x$-axis is down, and the $y$-axis is left.  We assume that the filament $(\mathbf{L})$ lies in the $x$--$z$ plane, making angle $\theta_L$ to the line-of-sight.  Equivalently, the directions can be expressed as a rotation around the $y$-axis:
\begin{equation}
\hat {\mathbf L} = {\mathbf R}_{\hat {\mathbf y}} (\theta_L) \hat {\mathbf z}.
\end{equation}
The magnetic field $(\mathbf{H})$ we describe with respect to the filament axis, using spherical-polar coordinates $(\theta_{LH},\phi_{LH})$:
\begin{equation}
\hat {\mathbf H} = {\mathbf R}_{\hat {\mathbf L}} (\phi_{LH}) {\mathbf R}_{\hat {\mathbf y}} (\theta_{LH}) \hat {\mathbf L} 
\end{equation}
When we numerically generate realizations of these directions, we use the Rodrigues rotation formula to rotate the vectors around the proper axis.  We use the distribution of $\theta_{LH}$ values to statistically characterize the misalignment in three dimensions between the filaments and their local magnetic fields.

The important quantities for computing the polarization of this filament are the angle that the magnetic field makes with the line-of-sight, expressed as:
\begin{equation}
  \cos \theta_{H} =  \hat {\mathbf H} \cdot \hat {\mathbf z}.
\end{equation}
When projected onto the plane of the sky, the filament and the field are separated by a misalignment angle $\psi_{LH}$, computed as:
\begin{equation}
\tan \psi_{LH} =  \hat H_y/ \hat H_x
\end{equation}
in the proper quadrant.  For dust, the polarization angle relative to the filament direction is $\psi_{\rm pol} = \psi_{LH} + \pi/2$.

For a particular filament angle $\theta_L$ and field--filament misalignment $\theta_{LH}$, changing the angle $\phi_{LH}$ rotates the field $\mathbf{H}$ to sweep out a cone around the filament direction.  We can use the fact that $p(\phi_{LH})$ is uniform on $[0,2\pi]$ to numerically accumulate the joint distribution of the field angle and projected field--filament misalignment:
\begin{equation}
  p(\theta_H, \psi_{LH} | \theta_{L}, \theta_{LH}).
\end{equation}
Because of the projections, this makes a loop of probability in the $(\theta_H, \psi_{LH})$ parameter space.  When the separation between the filament and field $\theta_{LH}$ is small, the loop centers tightly around $\theta_H = \theta_L$, $\psi_{LH} = 0$, and when $\theta_{LH}$ is larger, the loop is larger and more distorted.

\begin{figure}
  \begin{center}
    \includegraphics[width=0.75\columnwidth]{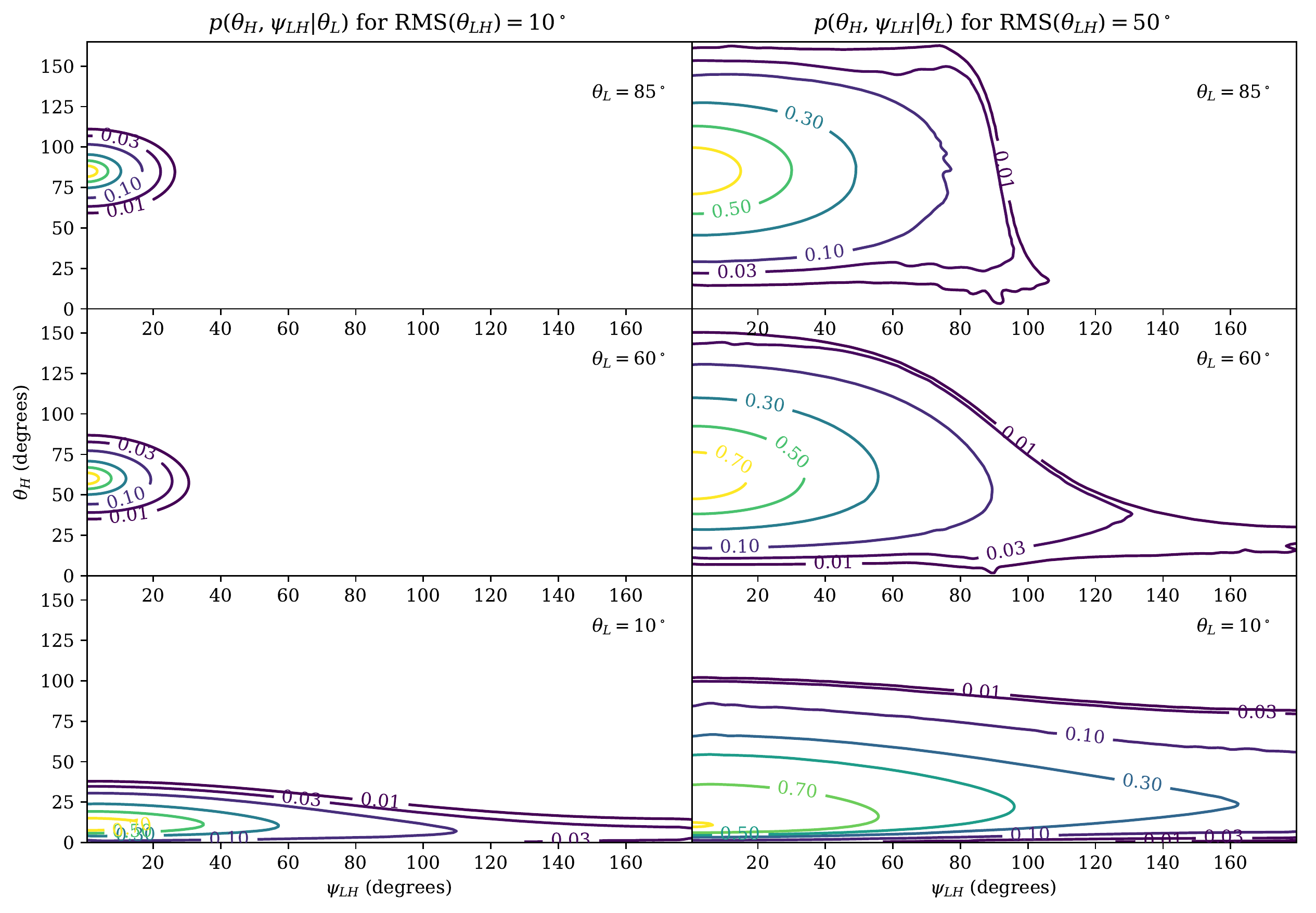}
  \end{center}
  \caption{The joint distribution of the line-of-sight angle of the magnetic field ($\theta_H$) and the projected angle between the magnetic field and the long-axis of the filament ($\psi_{LH}$), under the assumption that the magnetic field direction has a Gaussian  random distribution around the filament direction.  The contours mark lines of constant probability density, and the number records the integrated probability outside the contour.   The peak of the distribution is at $\psi_{LH}=0^\circ$, $\theta_H = \theta_L$, corresponding to a filament aligned with the local magnetic field.   The distribution is symmetric in the projected separation so we only show the half with $\psi_{LH} > 0^\circ$.  In the left column, the field and filament are more closely aligned (${\rm RMS}(\theta_{LH}) = 10^\circ$) than in the right column  (${\rm RMS}(\theta_{LH}) = 50^\circ$).  In the top row, the filament is perpendicular to the line-of-sight ($\theta_L = 90^\circ$) and so is in the plane of the sky.  In the bottom row, the filament aligns nearly along the line of sight ($\theta_L = 10^\circ$).    For filaments along the line of sight (small $\theta_L$), even a small misalignment with magnetic field can cause the projected angle ($\psi_{LH}$) to vary widely, and so the distribution of the projected angle is broad.
   }\label{fig:psiLH_thetaH_dist}
\end{figure}

\begin{figure}
  \begin{center}
    \includegraphics[width=0.49\columnwidth]{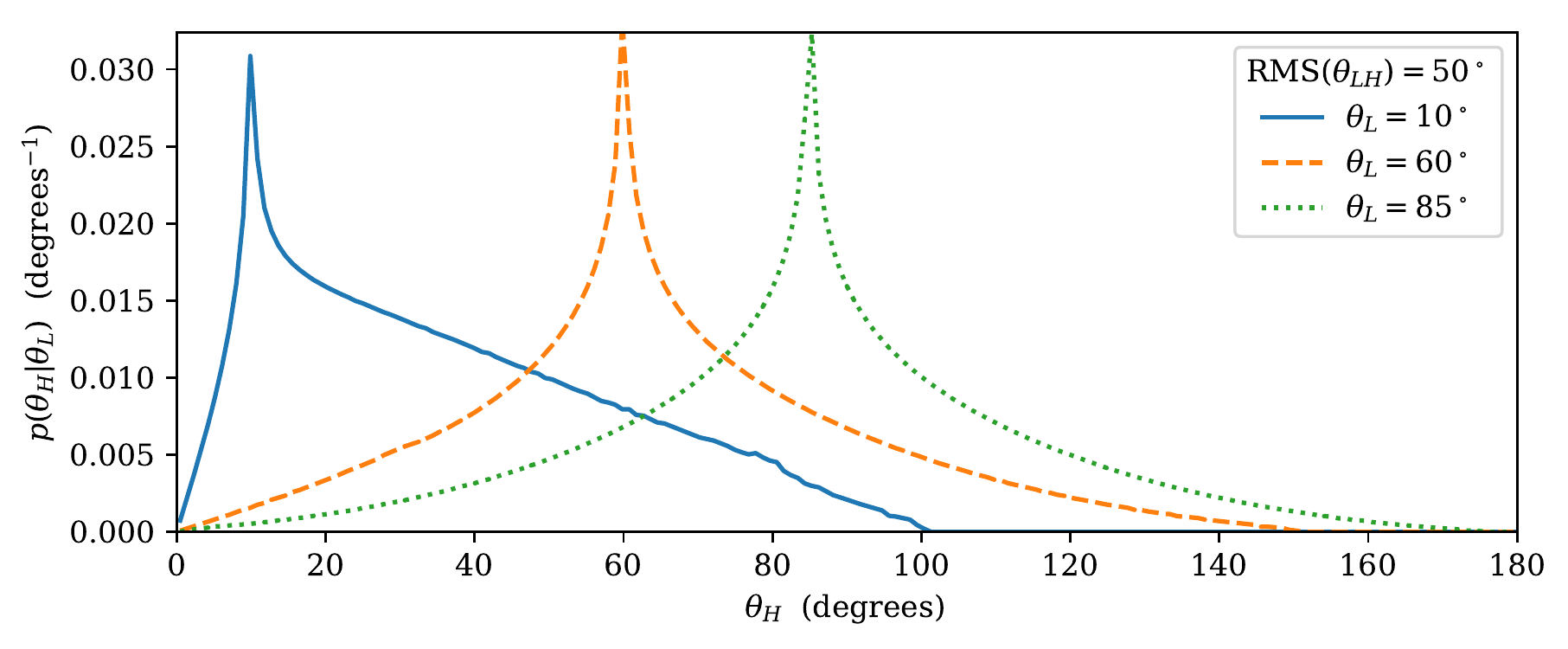}
    \hfill
    \includegraphics[width=0.49\columnwidth]{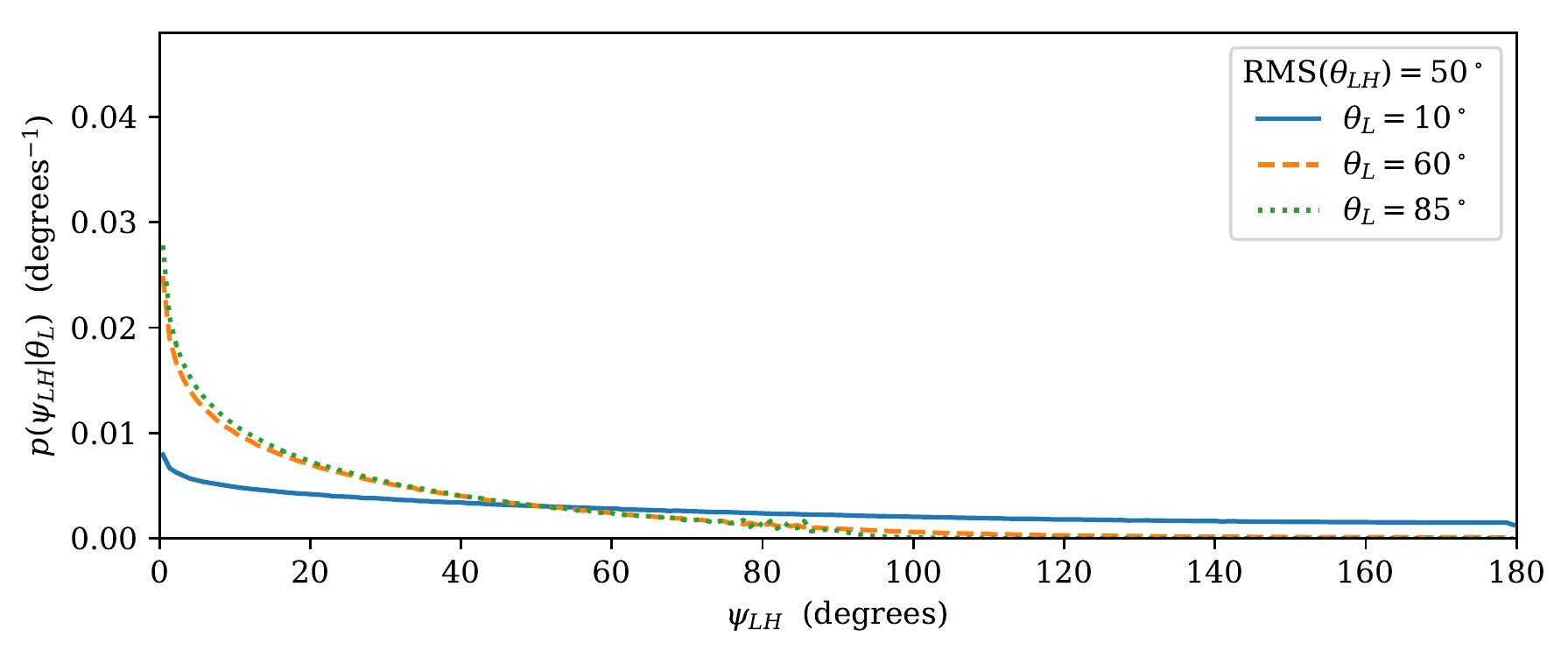}
  \end{center}
  \caption{(Left) Distribution of the magnetic field angle, for three different filament angles, assuming a Gaussian distribution for the misalignment of the field and filament angle in three-dimensions.  Each distribution peaks at the filament direction (for aligned filaments).  (Right) Distribution of the projected misalignment between the filament and the magnetic field.  The distribution is symmetric about $\psi_{LH} = 0^\circ$.}\label{fig:marginal_psiLH_thetaH_dist}
\end{figure}

\begin{figure}
  \includegraphics{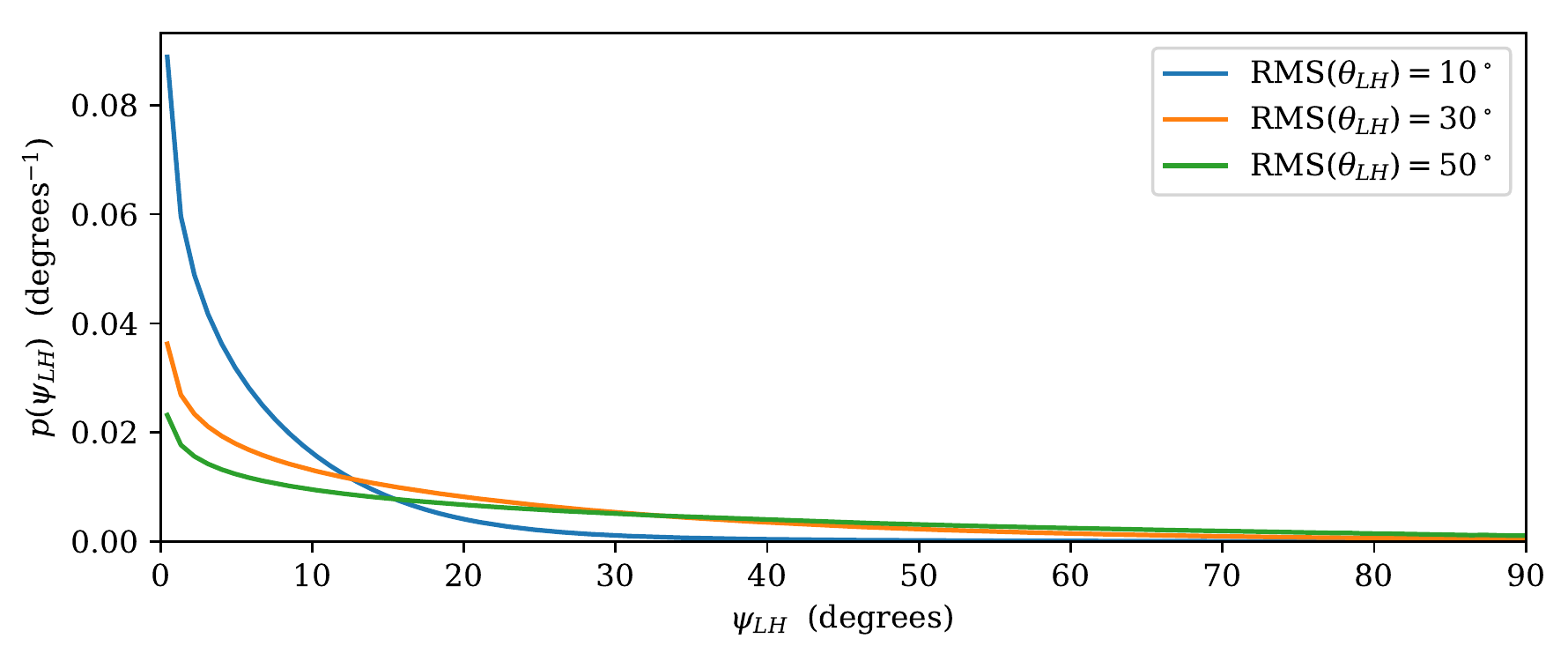}
\caption{The distribution of the projected filament--field misalignment angle for randomly oriented filaments, under the assumption that the three-dimensional misalignment angle is Gaussian distributed, for various misalignment dispersions.  The projected distribution is symmetric around $\psi_{LH} = 0$.}\label{fig:psiLH_marg}
\end{figure}

If we make an assumption for the distribution of the field--filament misalignment angle $\theta_{LH}$, we can marginalize over it.
\begin{equation}
  p(\theta_H, \psi_{LH} | \theta_L) = \int d\theta_{LH}\, p(\theta_H, \psi_{LH} | \theta_{L}, \theta_{LH}) p(\theta_{LH}) .
\end{equation} \label{eq:thetaH_psiLH}
In Fig.~\ref{fig:psiLH_thetaH_dist}, we show this joint distribution for several filament directions, under the assumption that the misalignment angle is Gaussian distributed.  For filaments along the line of sight (small $\theta_L$), the distribution in projected angle ($\psi_{LH}$) is broad, which makes sense as variations in the angle $\phi_{LH}$ cause a wide variety of $\psi_{LH}$ angles.
In Fig.~\ref{fig:marginal_psiLH_thetaH_dist}, we also examine the marginal distributions $p(\psi_{LH}|\theta_L)$ and $p(\theta_{H}|\theta_L)$.\footnote{At the beginning we fixed the plane-of-sky orientation of the filament, but we could have favored the magnetic field instead, and by symmetry we should have $p(\theta_{H}|\theta_L) = p(\theta_{L}|\theta_H)$.}
In Fig.~\ref{fig:psiLH_marg}, we further marginalize over the filament  angles to get the distribution of the projected field--filament misalignment:
\begin{equation}
  p(\psi_{LH}) = \int d\theta_{L}\, d\theta_{H}\,  p( \psi_{LH} | \theta_L)  p(\theta_{L}).
\end{equation}
These marginalized distribution have positive kurtosis and are more sharply peaked than the Gaussian distribution from which they are derived.
It is important to note that the polarized flux in practice will depend both on the polarization fraction (dependent on $\sin^2 \theta_H$) and the optical depth (dependent on the filament physical size, aspect ratio, and orientation $\theta_L$), and so the observed distribution of misalignment above some signal-to-noise cut will differ, and must be computed from the multidimensional distribution accounting for survey characteristics.

\end{document}